\documentclass[a4paper,12pt]{article}
\usepackage{bm, color}
\usepackage{amsmath,amsfonts,amssymb}
\newcommand{\ve}{\bm{\xi}}
\newcommand{\vp}{\bm{\xi'}}
\newcommand{\vs}{\bm{\xi}_{*}}
\newcommand{\vsp}{\bm{\xi'}_{*}}
\newcommand{\n}{\mathbf{n}}
\newcommand{\bV}{\mathbf{V}}
\newcommand{\bx}{\mathbf{x}}

\newcommand{\cE}{\mathit{E}}

\newcommand{\rea}{\mathbb{R}}
\newcommand{\irn}[1]{\int_{\rea^{#1}}}
\newcommand{\p}{\, .}
\newcommand{\sv}{\, ,}
\newcommand{\eq}[1]{(\ref{#1})}
\newcommand{\dm}{\displaystyle}

\newcommand{\setve}{\Omega}
\author{Armando Majorana
\\[5pt] Dipartimento di Matematica e Informatica
\\ Viale A. Doria 6, 95125 Catania, Italy}
\title{}
\title{A new macroscopic model derived from the Boltzmann equation and the discontinuous
Galerkin method for solving kinetic equations.}
\begin{document}
\maketitle
%
%
\begin{abstract}
We propose a new macroscopic model derived from the classical nonlinear Boltzmann equation.
A set of partial differential equations is obtained easily. 
The $5 N$ unknowns $N_{\alpha,j}(t, \bx)$  depend on the time and space coordinates, 
$\alpha = 1, 2, . ... N$ and $j = 0, 1 .., 4$. The index $\alpha$ labels a subset
$C_{\alpha}$ of a partition of the velocity space and the five indexes $j$ correspond to the
collision invariants.
The unknowns are related to the distribution function $f(t,\bx, \ve)$, solution of the
Boltzmann equation, since
\begin{eqnarray*}
&&
N_{\alpha,0}(t, \bx) \approx  \int_{C_{\alpha}} f(t,\bx, \ve) \: d \ve \sv
\qquad
\left(
\begin{array}{c}
N_{\alpha,1}(t, \bx) \\ N_{\alpha,2}(t, \bx) \\ N_{\alpha,3}(t, \bx)
\end{array}
\right)
\approx  \int_{C_{\alpha}} f(t,\bx, \ve) \, \ve \: d \ve \sv
\\
&&
N_{\alpha,4}(t, \bx) \approx \int_{C_{\alpha}} f(t,\bx, \ve) \, |\ve|^{2}\: d \ve \p
\end{eqnarray*}
This new model guarantees the conservation of the mass, momentum and energy.
We prove that the set of equations coincides with the set obtained applying the discontinuous
Galerkin method to the Boltzmamn equation \cite{AM}.
\end{abstract}
MSC-class: 76P, 82C40 (Primary) 65M60 (Secondary)
%
%
%
%
%
\clearpage
\section{Introduction and basic equations}
We consider the classical nonlinear Boltzmann equation  \cite{C88}, \cite{C90} for neutral
monatomic gases
\begin{equation}
\frac{\partial f}{\partial t} + \ve \cdot \frac{\partial f}{\partial \bx} =
Q(f,f)  \p \label{eqb}
\end{equation}
The one-particle distribution function $f$ depends on time $t$, position $\bm{x}$ and
velocity $\ve$.
We denote by $X \subseteq \rea^{3}$ the domain of the spatial coordinates $\bx$. The velocity
space is $\rea^{3}$.
The collision operator is defined as follows
\begin{equation}
 Q(f,f) = \irn{3} \irn{3} \irn{3}  W(\ve, \vs | \vp, \vsp) 
\left( f' f'_{*} - f f_{*} \right)  d \vs \, d \vp \, d \vsp 
\end{equation}
Here, as in the following, to simplify the notation, often we omit to write the variables $t$
and $\bx$, explicitly.
Moreover, it is here understood that
$$
 f' = f(t,\bx, \vp) , \quad
 f'_{*} = f(t,\bx, \vsp) , \quad
 f = f(t,\bx, \ve) , \quad
 f_{*} = f(t,\bx, \vs) \p
$$
The kernel $W$ of the collision operator is defined by
\begin{equation}
W(\ve, \vs | \vp, \vsp) = K(\n \cdot \bV, |\bV|) \,
\delta(\ve + \vs - \vp - \vsp) \,
\delta(|\ve|^{2} + |\vs|^{2} - |\vp|^{2} - |\vsp|^{2})
 \label{Wv4}  
\end{equation}
where
\begin{equation}
\n = \dfrac{\ve - \vp}{|\ve - \vp|}  \quad \mbox{and} \quad 
\bV = \ve - \vs \p \label{nV}
\end{equation}
The function $K$ is related to the interaction law between colliding particles. 
The Dirac distributions guarantee momentum and energy conservation during the binary
collisions.
The collision operator $Q(f,f)$ of the Boltzmann equation \eq{eqb} is usually written in
different way, since Dirac distributions are used to reduce the collision integral to a
five-fold integral.

Let $\phi : \rea^{3}  \rightarrow \rea$ be a measurable function. 
Now, we assume that generic test functions $\phi$ depend only on the velocity $\ve$.
If we multiply both sides of the Boltzmann equation \eq{eqb} by a test function $\phi(\ve)$
and we integrate with respect to the velocity $\ve$, then we obtain the equation
\begin{equation}
 \frac{\partial \mbox{ }}{\partial t} \irn{3} f(t,\bx, \ve) \, \phi(\ve) \: d \ve 
 + \nabla_{\bx} \! \irn{3}  f(t,\bx, \ve) \, \ve \, \phi(\ve) \: d \ve =
\irn{3} Q(f,f)(t,\bx, \ve) \, \phi(\ve) \: d \ve \p \label{eqBve}
\end{equation}
For any function $\phi$, we obtain an equation of type \eq{eqBve}.
This procedure recalls the classical moment method, but, in our framework, the test functions
will have compact support.

We recall a well-known result.
Assuming the existence of the integrals, the right hand side of Eq.~\eq{eqBve} 
can be written as follows
\begin{equation}
\dfrac{1}{2} \irn{3} \irn{3} \irn{3} \irn{3}   W(\ve, \vs | \vp, \vsp)
\left[ \phi' + \phi'_{*} - \phi - \phi_{*} \right] 
f \,  f_{*} \: d \vp \, d \vsp d \vs \, d \ve \p 
\label{Qphi}
\end{equation}
Now, if we define
\begin{equation}
\mathcal{K}(\phi; \ve , \vs) = \irn{3} \irn{3}  W(\ve, \vs | \vp, \vsp)
\left[ \phi(\vp) + \phi(\vsp) -  \phi(\ve) - \phi(\vs) \right] d \vp \, d \vsp
\label{Kphi}
\end{equation}
then the integral \eq{Qphi} becomes
\begin{equation}
\irn{3} Q(f,f) \, \phi(\ve) \: d \ve  =
\frac{1}{2} \irn{3} \irn{3} \mathcal{K}(\phi; \ve , \vs) \,
f(\ve) \,  f(\vs) \: d \vs \, d \ve 
\label{eqQK}
\end{equation}
and Eq.~\eq{eqBve} writes
\begin{eqnarray}
 \frac{\partial \mbox{ }}{\partial t} \irn{3} f(t,\bx, \ve) \, \phi(\ve) \: d \ve 
 + \nabla_{\bx} \! \irn{3}  f(t,\bx, \ve) \, \ve \, \phi(\ve) \: d \ve 
& & \nonumber 
\\[7pt]
\mbox{} =
\frac{1}{2} \irn{3} \irn{3} \mathcal{K}(\phi; \ve , \vs) \,
f(t,\bx,\ve) \,  f(t,\bx,\vs) \: d \vs \, d \ve \p 
& & \label{BKphi}
\end{eqnarray}
We note that the function $\mathcal{K}(\phi; \ve , \vs)$, which is related to the function
$\phi$, depends only on the $(\ve, \vs)$ variables and plays the role of a kernel of the
integral operator \eq{eqQK}.
We recall a simple but important result.
If $\psi$ denotes one of the collision invariants $1$, $\ve$, $| \ve |^{2}$, then we have
\begin{equation}
\mathcal{K}(\psi; \ve , \vs)  = 0
\quad \forall \, \ve , \vs \in \rea^{3} . \label{Kpsi} 
\end{equation}
We denote by $\left\lbrace \psi_{j} : \: j=0,1..,4 \right\rbrace$ the ordered set
$\left\lbrace 1, \xi_{1}, \xi_{2}, \xi_{3}, |\ve|^{2} \right\rbrace$,
where $\xi_{1}, \xi_{2}, \xi_{3}$ are the three components of the vector $\ve$.
%
%
%
%
%
%
\section{The macroscopic model}
As in the numerical treatment of a kinetic equation by means of finite differences or
elements, we requires a bounded domain for the velocity space. To this scope, we introduce a
suitable characteristic function in the kernel of the collision operator, such that there
exists a bounded domain $\setve \subseteq \rea^{3}$ so that, if $f(0,\bx,\ve) = 0$ for every
$\ve \not \in \setve$ and $\bx \in X$, then $f(t,\bx,\ve) = 0$ for every $\ve \not \in \setve$,
$\bx \in X$  and for all time $t$.
Let $\cE$ be a suitable positive real number.
We define the function $ \chi_{\cE} : \rea^{3} \times \rea^{3} \rightarrow \rea$ as
follows
$$
 \chi_{\cE}(\ve, \vs) = 
\left\lbrace 
\begin{array}{ll}
1 &  \mbox{if } |\ve|^{2} + |\vs|^{2} \leq \cE \\[5pt]
0 & \mbox{otherwise}
\end{array}
\right. \sv
$$
and
\begin{equation}
 W_{\cE}(\ve, \vs | \vp, \vsp) = \chi_{\cE}(\ve, \vs) \, W(\ve, \vs | \vp, \vsp)
 \p
\end{equation}
The modified kernel $W_{\cE}$ has the same fundamental properties of the true kernel $W$
and guarantees that if two particles have velocities $\ve$ and $\vs$ before the impact,
such that $|\ve|^{2} + |\vs|^{2} \leq \cE$, then the velocities $\vp$ and $\vsp$ of the
particles, after the impact, will satisfy the inequality $|\vp|^{2} + |\vsp|^{2} \leq \cE$.
We can choose $\setve = \left\lbrace \ve \in \rea^{3} \: : |\ve|^{2} \leq \cE \right\rbrace$,
and we consider the Boltzmann equation \emph{with the modified kernel $W_{\cE}$}.

Now, we choose $N$ measurable subsets $C_{\alpha}$ $(\alpha = 1, 2, ,.., N)$ of $\setve$
such that
$$
C_{\alpha} \subseteq \setve \quad \forall \alpha \sv \quad
C_{\alpha} \cap C_{\beta} = \emptyset 
\quad \forall \, \alpha \neq \beta \quad \mbox{and} \quad
\bigcup_{\alpha = 1}^{N} C_{\alpha} = \setve \p
$$
We remark that we have not introduced any constrain on the size and shape of the cells;
so, there is a great arbitrariness in the decomposition of the domain $\setve$.  
We denote by $\chi_{\alpha}$ the characteristic function on the set $C_{\alpha}$.
\\
\emph{We consider the $5N$ test functions}
\begin{equation}
\psi(\ve) \, \chi_{\alpha}(\ve) \quad \mbox{ for } 
\psi(\ve) = 1, \xi_{1}, \xi_{2}, \xi_{3}, |\ve|^{2} \mbox{ and }
\alpha = 1, 2, . ... N \p
\end{equation} 
Therefore, Eq.~\eq{BKphi} furnishes the $5N$ equations
\begin{eqnarray}
& &
 \frac{\partial \mbox{ }}{\partial t} \int_{C_{\alpha}} f(t,\bx, \ve) \, \psi_{j}(\ve) 
 \: d \ve 
 + \nabla_{\bx} \int_{C_{\alpha}}  f(t,\bx, \ve) \, \ve \, \psi_{j}(\ve) \: d \ve 
\nonumber 
\\[7pt]
& & \mbox{} =
\frac{1}{2} \int_{\setve} \int_{\setve} \chi_{\cE}(\ve, \vs) \, 
\mathcal{K}(\psi_{j} \, \chi_{\alpha} ; \ve , \vs) \,
f(t,\bx,\ve) \,  f(t,\bx,\vs) \: d \vs \, d \ve \sv \label{eqfa}
\end{eqnarray}
where $\alpha = 1, 2, . ... N$ and $j = 0, 1 .., 4$.
\\
We choose the $5 N$ scalar integrals 
\begin{equation}
N_{\alpha, j}(t, \bx) =
 \int_{C_{\alpha}} f(t,\bx, \ve) \, \psi_{j}(\ve) \: d \ve \label{faphi} 
\end{equation}
as the only independent variables and we look for a reasonable closure of the system \eq{eqfa}.
\\
The physical meaning of the new variables is evident; for instance, $N_{\alpha, 0}(t, \bx)$ is
the density of particles at time $t$ and position $\bx$ having velocity belonging to the set
$C_{\alpha}$.

A very simple recipe for a closure is the following.
\\
Firstly, we treat the drift term of Eq.~\eq{eqfa}.
We consider a generic cell $C_{\alpha}$ and we look for an approximation in $C_{\alpha}$ of the
function $\ve \, \psi_{j}(\ve)$ $(j = 0, 1 .., 4)$ of this type
\begin{equation}
 \xi_{i} \, \psi_{j}(\ve) \approx 
\sum_{k=0}^{4} \left[ \mathbf{a}^{(\alpha)}_{kj} \right]_{i} \psi_{k}(\ve)
\quad \mbox{for every } \ve \in C_{\alpha} \mbox{ and } i=1,2,3 \sv
\label{coef_a}
\end{equation}
where 
$ \dm \mathbf{a}^{(\alpha)}_{kj} = \left( 
\left[ \mathbf{a}^{(\alpha)}_{kj} \right]_{1} \sv
\left[ \mathbf{a}^{(\alpha)}_{kj} \right]_{2} \sv
\left[ \mathbf{a}^{(\alpha)}_{kj} \right]_{3} \,
\right)$ 
is a constant array to be determined.
\\[5pt]
Now, we consider the collision term of Eq.~\eq{eqfa}.
Since
\begin{eqnarray*}
& &
\frac{1}{2} \int_{\setve} \int_{\setve} \chi_{\cE}(\ve, \vs) \, 
\mathcal{K}(\psi_{j} \, \chi_{\alpha} ; \ve , \vs) \,
f(t,\bx,\ve) \,  f(t,\bx,\vs) \: d \vs \, d \ve
\\[7pt]
& & \mbox{} =
\frac{1}{2} \sum_{\beta=1}^{N} \sum_{\gamma=1}^{N}
\int_{C_{\beta}} \int_{C_{\gamma}} \chi_{\cE}(\ve, \vs) \, 
\mathcal{K}(\psi_{j} \, \chi_{\alpha} ; \ve , \vs) \,
f(t,\bx,\ve) \,  f(t,\bx,\vs) \: d \vs \, d \ve \sv 
\end{eqnarray*}
we look for an approximation of this type
\begin{equation}
\chi_{\cE}(\ve, \vs) \, \mathcal{K}(\psi_{j} \, \chi_{\alpha} ; \ve , \vs) 
\approx
\sum_{k=0}^{4} \sum_{n=0}^{4} b^{(\alpha \beta \gamma)}_{jkn}
\psi_{k}(\ve) \, \psi_{n}(\vs)
\quad \mbox{for every } (\ve, \vs) \in C_{\beta} \times C_{\gamma} \p
\label{coef_b}
\end{equation}
Here, $b^{(\alpha \beta \gamma)}_{jkn}$ are the numerical parameters to be determined.
\\
If we assume reasonable the approximations given by Eqs.~\eq{coef_a}-\eq{coef_b}, we obtain
the following closed system of partial differential equations
\begin{equation}
\frac{\partial N_{\alpha,j}}{\partial t} +
\sum_{k=0}^{4} \mathbf{a}^{(\alpha)}_{kj} \cdot \nabla_{\bx} N_{\alpha,k} =
\frac{1}{2} \sum_{\beta=1}^{N} \sum_{\gamma=1}^{N}
\sum_{k=0}^{4} \sum_{n=0}^{4} b^{(\alpha \beta \gamma)}_{jkn}
N_{\beta,k} \, N_{\gamma,n} \p \label{eq_macro}
\end{equation}
At this point, we must give a meaning to the approximations given by
Eqs.~\eq{coef_a}-\eq{coef_b}.
\\
For each cell $C_{\alpha}$ we introduce a vector function $\bm{\eta}_{\alpha}(\ve)$.
The components of the five dimensional array $\bm{\eta}_{\alpha}(\ve)$ are functions,
denoted by $\eta_{\alpha, i}(\ve)$, which are linear combination of the collision
invariants and such that
\begin{equation}
\int_{C_{\alpha}} \eta_{\alpha, i}(\ve) \, \psi_{j}(\ve) \: d \ve = \delta_{i j} 
\label{orto}
\end{equation}
for every $i$ and $j$ and for each cell $C_{\alpha}$.
\\
We determine the numerical parameters $\mathbf{a}^{(\alpha)}_{kj}$ and 
$b^{(\alpha \beta \gamma)}_{jkn} $, for every
$(\alpha , \beta , \gamma = 1, ..., N)$ and $(j, k, n = 0, ..., 4)$,
assuming that the following equations hold
\begin{eqnarray}
&&
\int_{C_{\alpha}} \eta_{\alpha, i}(\ve) \, \ve \, \psi_{j}(\ve) \: d \ve
=
\sum_{k=0}^{4} \mathbf{a}^{(\alpha)}_{kj} 
\int_{C_{\alpha}} \eta_{\alpha, i}(\ve) \, \psi_{k}(\ve)  \: d \ve \sv
\label{eq_a}
\\
&&
\int_{C_{\beta}} \int_{C_{\gamma}} \eta_{\beta, i}(\ve) \, \eta_{\gamma, m}(\vs) \,
\chi_{\cE}(\ve, \vs) \, \mathcal{K}(\psi_{j} \, \chi_{\alpha} ; \ve , \vs) 
\: d \vs \, d \ve \nonumber
\\
&& \mbox{} =
\sum_{k=0}^{4}  \sum_{n=0}^{4} b^{(\alpha \beta \gamma)}_{jkn}
\int_{C_{\beta}} \int_{C_{\gamma}} \eta_{\beta, i}(\ve) \, \eta_{\gamma, m}(\vs) \,
\psi_{k}(\ve) \, \psi_{n}(\vs) \: d \vs \, d \ve \p
\label{eq_b}
\end{eqnarray}
Taking into account Eq.~\eq{orto}, we obtain
\begin{eqnarray}
&&
\mathbf{a}^{(\alpha)}_{ij} =
\int_{C_{\alpha}} \eta_{\alpha, i}(\ve) \, \ve \, \psi_{j}(\ve) \: d \ve \sv
\\
&&
b^{(\alpha \beta \gamma)}_{jim} =
\int_{C_{\beta}} \int_{C_{\gamma}} \eta_{\beta, i}(\ve) \, \eta_{\gamma, m}(\vs) \,
\chi_{\cE}(\ve, \vs) \, \mathcal{K}(\psi_{j} \, \chi_{\alpha} ; \ve , \vs) 
\: d \vs \, d \ve \p
\end{eqnarray}
We note that the parameters 
$\mathbf{a}^{(\alpha)}_{kj}$ and $b^{(\alpha \beta \gamma)}_{jkn} $
are \emph{numerical constants}, which depend only on the domain decomposition and the
scattering kernel $K$. Therefore, they do not depend on the solutions of the Bolzmann equation.
\\
It is immediate to verify that, with this choice of the parameters, the macroscopic equations
\eq{eq_macro} coincide with the equations (Ref. \cite{AM}) obtained applying the discontinuous
Galerkin method to the Boltzmann equation \eq{eqb}.
Moreover, as a consequence of a result of paper \cite{AM}, it is guaranteed the conservation of
mass, momentum and energy for homogeneous solutions.
%
%
%
%
%
%

%
\end{document}